# The Effects of Website Quality on Adoption of E-Government Service: An Empirical Study Applying UTAUT Model Using SEM

Mohammed Alshehri
School of Information & Communication Technology
Griffith University
Brisbane, Australia
E-mail: m.alshehri@griffith.edu.au

Steve Drew
School of Information & Communication Technology
Griffith University
Brisbane, Australia
E-mail: s.drew@griffith.edu.au

Thamer Alhussain
Computer Sciences & Information Technology College
King Faisal University
Ahsaa, Kingdom of Saudi Arabia
E-mail: talhussain@kfu.edu.sa

Rayed Alghamdi
Faculty of Computing & Information Technology
King Abdulaziz University
Jeddah, Kingdom of Saudi Arabia
E-mail: raalghamdi8@kau.edu.sa

## Abstract

*In today's global age, e-government services have become the main channel for online communication between the government and its citizens. They aim to provide citizens with more accessible, accurate, real-time and high quality services. Therefore, the quality of government websites which provide e-services is an essential factor in the successful adoption of e-government services by the public. This paper discusses an investigation of the effect of the Website Quality (WQ) factor on the acceptance of using e-government services (G2C) in the Kingdom of Saudi Arabia (KSA) by adopting the Unified Theory of Acceptance and Use of Technology (UTAUT) Model. Survey Data collected from 400 respondents were examined using the structural equation modelling (SEM) technique and utilising AMOS tools. This study found that the factors that significantly influenced the Use Behaviour of e-government services in KSA (USE) include Performance Expectancy (PE), Effort expectancy (EE), Facilitating Conditions (FC) and Website Quality (WQ), while the construct known Social Influence (SI) did not. Moreover, the results confirm the importance of quality government websites and support systems as one of the main significant and influential factors of e-government services adoption. The results of this study can be helpful to Saudi's governmental sectors to adjust their corporate strategies and plans to advance successful adoption and diffusion of e-government services (G2C) in KSA.*

## INTRODUCTION

In the current era of information technology and Internet services, governments around the world started their e-government services systems to provide government services and information to their citizens in a professional manner that is also secure, safe and offers to save time for its users. E-government services aim to improve the quality and assurance of government services and allow for time and effort savings in government administration. According to Carter and Belanger (2005), e-government services increase the convenience and accessibility of government services and information to citizens. Citizens require government to provide e-services replete with high quality, quantity, and constant availability to fulfil the demands their citizens. Carter and Belanger (2003, 2005), Pavlou (2003), and Gefen (et al. 2003) all reported that one of the most important factors for the success of e-government services (G2C) adoption is citizens' acceptance and use of those services. Moreover, the quality of government websites has become a key indicator of citizens' satisfaction and acceptance of e-government services. However, most e-government websites are developed by IT staff and experts in government sectors who do not consider citizens' perceptions or needs; naturally, this creates obstacles to winning user acceptance of e-government services. In addition, Al-Nuaim's (2011) study assessed the current state of the Saudi e-government by evaluating its ministries' websites using a citizen-centred e-government approach. It was found that 8 out of 21 (41%) ministries did not implement the main features of an e-government web site. In addition, 10 ministries (45.4%) were completely or partially in the first stage (web presence); 3 ministries (13.6%) were in the second stage (one-way interaction); and 6 ministries had no online service at all. Although Saudi Arabia has the largest and fastest growing ICT marketplace in the Arab region, government online services have not progressed at a similar speed (AlGhamdi et al. 2011). Therefore, this paper addresses the issue of e-government website quality and aims to:

1. Understand the citizen's perception of e-government website quality;
2. Develop and test the UTAUT model that captures the website quality construct based on quantitative data; and
3. Investigate the relationship between website quality and the behavioural intentions of citizens who use or could use e-government services.

This paper is structured as follows. First, the e-government background and the importance of website quality are presented. The technology adoption models and the UTAUT model are elaborated in the second section. Then, in the third section, the research methodology which includes the research model, hypotheses, data collection and analysis are stressed. In the fourth section, the findings and results are presented, including an evaluation of the measurement and structural model. Fifth and finally, a discussion of the findings and an agenda for future research are provided.

## E-GOVERNMENT: BACKGROUND

E-government represents an essential change in the whole public sector structure, values, culture and ways of conducting business. In fact, there are many definitions for the term e-government and the differences reflect the varying priorities in government strategies. Moon and Norris (2005) provides a simple definition; e-government is perceived as a "means of delivering government information and service" (p. 43). Isaac (2007) defined electronic government as the government's use of technology, particularly web-based Internet applications, to enhance the access to and delivery of government information and service to citizens, business partners, employees, other agencies, and government entities. Fang (2002) defined e-government as a way for governments to use the most innovative information and communication technologies, particularly web-based Internet applications, to provide citizens and businesses with more convenient access to government information and services, to improve the quality of the services, and to provide greater opportunities to participate in democratic institutions and processes. Further, Carter and Belanger (2005) defined e-government services as the use of ICT to enable and improve the efficiency of government services provided to citizens, employees, businesses, and agencies. According to Carter and Belanger (2005), e-government services increase the convenience and accessibility of government services and information to citizens. Nowadays, government agencies around the world are making their services increasingly available online. E-government services in this paper refer to all Government to Citizen (G2C) communication over Internet and Web-based applications which aims to facilitate, deliver and improve the quality of government services.

## WEBSITE QUALITY IN THE E-GOVERNMENT MODEL

Aladwani and Palvia (2002) defined web quality as a user's positive evaluation of a website's features, ensuring it meets the user's needs and reflects the overall excellence of the website. Therefore, they identified three dimensions of web quality: technical adequacy, web content, and web appearance. Moreover, Zhong and Ying (2008) stated that website quality includes the features of the website system which present measures of quality such as system, information, and service quality. In the website quality literature, several researchers have declared that website quality includes multiple dimensions, such as information quality, system quality, security, ease of use, user satisfaction, and service quality (Aladwani and Palvia 2002; DeLone and McLean 2003; Hoffman and Novak 2009; Urban et al. 2009). Furthermore, Floh and Treiblmaier (2006) emphasized that website quality, which include web design, structure and content, is an important factor for achieving customer satisfaction. Schaupp et al. (2006) conducted a survey to investigate the impact of information quality and system quality on website satisfaction. The results showed that information quality and system quality w e r e significant predictors of website satisfaction, and, therefore, intention to use the website. In addition, Li and Jiao (2008) confirmed that there is a significant relationship between website quality and user satisfaction and that this relationship affects the actual use of online services. In addition, website quality perceptions have been reported to affect behavioural intention and usage decisions in many studies, such as those by Ahn et al. (2007), Collier and Bienstock ( 2009), Nelson et al. (2005), Parasuraman et al. (2005), and Wixom and Todd (2005). It is clear that the quality of government websites which provide e-services is an essential factor and needs to be investigated and included in the proposed model. If e-government website design is of a professional standard with high quality, then it will promote user satisfaction and facilitate adoption.

## TECHNOLOGY ADOPTION MODELS

Technology acceptance is defined as "an individual's psychological state with regard to his or her voluntary or intended use of a particular technology" (Gattiker 1984).Technology acceptance models aim to study how to promote technology use and to explore the factors that hinder or facilitate the acceptance and use of technologies (Kripanont 2007). A number of technology models have developed over the years to study and investigate the effect of factors on the acceptance and use of technologies such as:

- Theory of Reasoned Action (TRA) (Ajzen and Fishbein 1980);
- Theory of Planned Behaviour (TPB) (Ajzen 1985);
- Technology Acceptance Model (TAM) (Davis 1989);
- Model of PC Utilization (MPCU)( Thompson et al. 1991);
- Motivational Model (MM) (Davis et al. 1992);
- Social Cognitive Theory (SCT) (Brown 1999);
- Extension of the Technology Acceptance Model (TAM2) (Venkatesh and Davis 2000);
- Diffusion of Innovation Model (DOI) (Rogers 2003); and the
- Unified Theory of Acceptance and Use of Technology (UTAUT) (Venkatesh et al. 2003).

In this study, the UTAUT model will be utilized to discover factors that influencing the adoption of e-government services in KSA where e-government services are still in the early development stages. The UTAUT model was empirically validated model and able to account for 70 percent of the variance in usage intention /behaviour. This result was a significant improvement over any of the eight original models where the maximum was around 40 percent. The UTAUT model will be discussed in detail in the following section.

### Unified Theory of Acceptance and Use of Technology (UTAUT)

The Unified Theory of Acceptance and Use of Technology (UTAUT) (Venkatesh et al. 2003) is one of the latest developments in the field of general technology acceptance models. Like earlier acceptance models, it aims to explain user intentions to use an IS and increase usage behaviour. Venkatesh et al. (2003) developed this synthesized model to present a more complete picture of the acceptance process than previous individual models had been able to do. The UTAUT model successfully integrates key elements from eight models previously used in the IS field. These models are the TRA, TPB, TAM, MPCU, MM, SCT, TAM2, and DOI and they each attempt to predict and explain user behaviour using a variety of independent variables. A unified model was created based on the conceptual and empirical similarities across these eight models. The theory holds that four key constructs (performance expectancy, effort expectancy, social influence, and facilitating conditions) are direct determinants of usage intention and behaviour (Venkatesh et al. 2003). The variables of gender, age, experience and voluntariness of use all work to moderate the impact of the four key constructs on usage intention and behaviour as indicated in Figure 1. The UTAUT has four core determinants that influence behavioural intention (BI) to use a technology; these determinants are defined as follows (Venkatesh et al. 2003, pp 447-453):

- Performance expectancy (PE): " the degree to which an individual believes that using the system will help him or her to attain gains in job performance";
- Effort expectancy (EE): "the degree of ease associated with use of the system";
- Social influence (SI): "the degree to which an individual perceives that important others believe he or she should use the new system"; and
- Facilitating conditions (FC): "the degree to which an individual believes that an organizational and technical infrastructure exists to support use of the system."

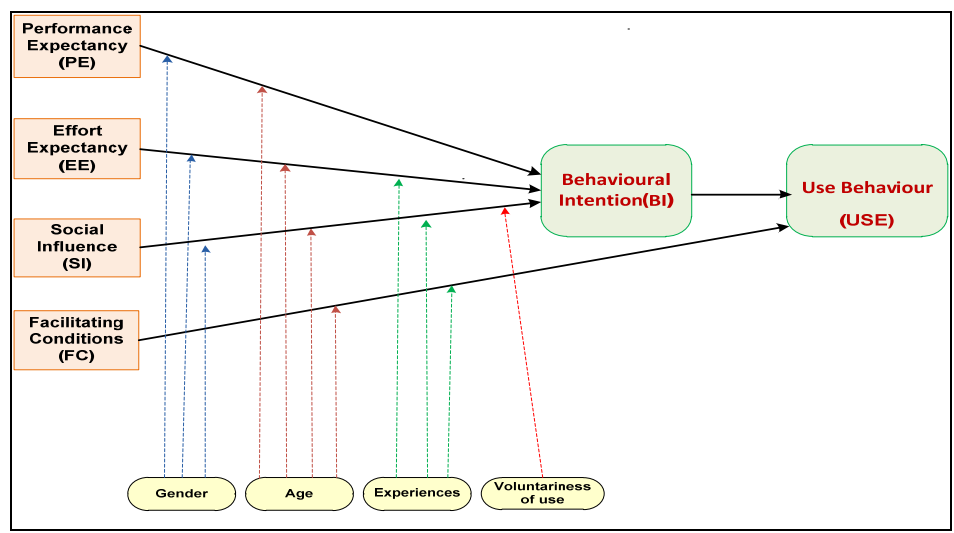

Figure 1: UTAUT model (Venkatesh et al. 2003)

## RESEARCH METHOD

### Research Model and Hypotheses

The objective of this research is to study the effect of website quality on the actual usage of e-government services from the perspective of Saudi Arabian citizens. The proposed model for this study is presented in Figure 2. It is mostly derived from the UTAUT model but with the following modifications:

- The experienced moderator in Venkatesh et al. (2003) was replaced with Internet experience. Several studies have shown that Internet experience influences both perceived usefulness and ease of use which, in consequence, affects people's actual use or intention to use specific systems (Agarwal and Prasad, 1999; Jiang et al. 2000). E-government services are more likely to be used by experienced Internet users.
- Website quality (WQ) has been added as an independent variable to the original UTAUT model and is moderated by gender, age and Internet experience. These variables will assist in understanding the influence of governments' website quality on citizens' perceptions and their willingness to adopt this new form of online services. Zhong and Ying (2008) stated that website quality (WQ) is the quality of the website itself or the services provided by that web system. Therefore, this definition of quality is based on two pillars: website quality and information quality. Website quality includes many features, such as website design, website functions, security, and information quality; these are measured by reliability, responsiveness, empathy, clarity and accuracy in the information and procedures (Ahn et al. 2003). In this study, five questionnaire statements are used to study the effect of website quality on actual usage of e-government service (see Appendix).
- A usage behaviour construct has been used to measure the actual usage of e-government services. Usage intention measures are particularly useful in the context of this study, as only the Saudi citizens who require and need to use e-government services are targeted. Therefore, in this case, an actual usage measure can lead to an accurate conclusion about the acceptability and suitability of e-government services.

However, in the interests of brevity for this paper, only the main hypotheses will be investigated, while the effect of the moderators will not be the focus of this study.

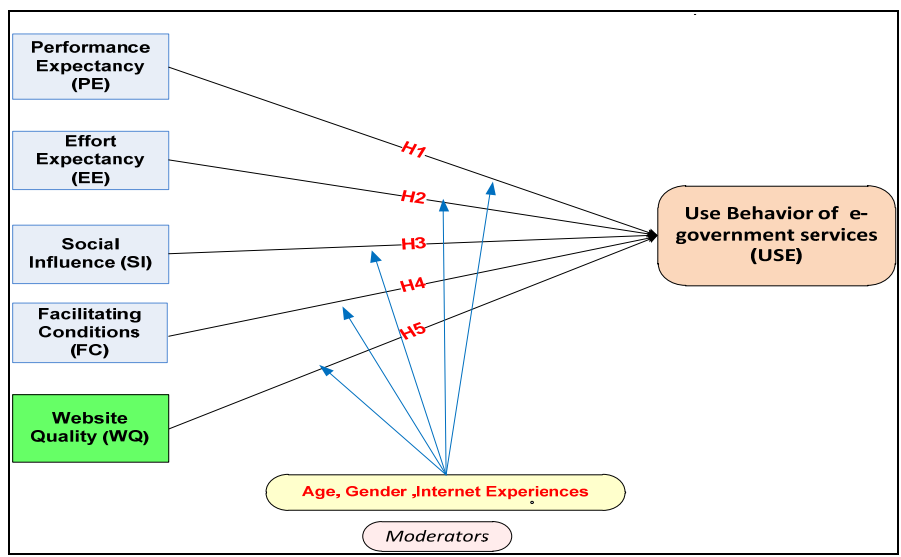

Figure 2: Research model

The researchers hypothesized relationships between variables as follows:

Table 1. Research Hypotheses

| No. | Hypotheses |
| --- | --- |
| H1 | Performance expectancy (PE) will have a positive and significant influence on usage behaviour to use e-government services (USE). |
| H2 | Effort expectancy (EE) will have a positive and significant influence on usage behaviour to use e-government services (USE). |
| H3 | Social influence (SI) will have a positive and significant influence on usage behaviour to use e-government services (USE). |
| H4 | Facilitating conditions (FC) will have a positive and significant influence on usage behaviour to use e-government services (USE). |
| H5 | Website quality (WQ) will have a positive and significant influence on usage behaviour to use e-government services (USE). |

## Data Collection

A quantitative questionnaire is employed in this study to collect empirical data. The questionnaire instrument is one of the most common tools of technology adoption as it uses a set of specific questions to cover the study topic and to target a large number of participants in a practical and efficient way (Carter and Belanger 2003, 2004; Reddick 2005; Venkatesh et al. 2003). The proposed UTAUT model includes five constructs and each construct is measured with multiple items. All UTAUT items were adapted from Venkatesh et al. (2003) while WQ items were adapted from Aladwani (2006) to improve content validity (Straub et al. 2004). The questionnaire items and their sources are listed in the Appendix. The sample for this study consists of Saudi citizens and the research questionnaires were distributed to 600 participants randomly chosen from three big cities in different geographic regions: Riyadh, Jeddah and Abha. This broad selection allowed for the inclusion of a culturally diverse range of citizens as subjects for the study. A total of 400 (66.6%) of questionnaires were returned and considered usable to analyse and fulfil the aim of this study.

## Data Analysis

Having completed the research method and data collection requirements, the next step is data analysis method and techniques. Data were analysed using the Structural Equation Modelling (SEM) approach and utilized AMOS tools. The SEM technique has been employed in this research to evaluate the relationships between the model constructs. Also, SEM has been used to model the complex relationship of multiple independent and dependent constructs (Kline 2005). Gefen et al. (2000) highly recommend the use of SEM in both behavioural sciences and IT/IS research. This study follows the two-step approach recommended by Anderson and Gerbing (1988); first, the measurement model was assessed to examine reliability and validity and, second, the structural model was assessed to test the research hypotheses and the suitability of the model.

## FINDINGS AND RESULTS

### Sample Profile

Table 2, presented below, provides a general demographic overview of the Saudi citizens who participated in this study in terms of gender, age and education level.

Table 2. Demographic Information of Respondents

| Variable | | Frequency | Percent |
|---|---|---|---|
| Gender | Male | 290 | 72.5 |
| | Female | 110 | 27.5 |
| Age | Less than 20 | 55 | 13.7 |
| | 21-30 | 195 | 48.7 |
| | 31-40 | 132 | 33.2 |
| | 41-50 | 15 | 3.7 |
| | More than 50 | 3 | 0.7 |
| Education | High School | 22 | 5.5 |
| | Diploma | 168 | 42.0 |
| | Bachelor | 195 | 48.7 |
| | Higher education | 15 | 3.8 |

**Evaluation of the Measurement Model**

The measurement model (a CFA model) specifies the relationships that suggest how measured variables represent a construct that is not measured directly (Hair et al. 2006). It was assessed with confirmatory factor analysis (CFA) using the AMOS 19.0 tool to examine convergent and discriminant validity. In the confirmatory factor analysis, the convergent validity relied on three indicators: the item reliability of each measure (factor loading); the reliability of each construct; and the average variance extracted (AVE). Constructs have convergent validity when the composite reliability exceeds the criterion of 0.70 and the average variance extracted (AVE) is above 0.50 (Hair et al. 2006). Table 3 shows the factor loadings, the average variance extracted (AVE), the composite reliability (CR), and the Cronbach Alpha values. All item loadings are bigger than 0.7 and *t* values indicate that all loadings are significant at 0.001. All AVEs were above 0.5 and all CRs were above 0.7. Therefore, the results support the convergent validity of the scales (Gefen et al. 2000; Hair et al. 2006). In addition, all Alpha values are larger than 0.7, revealing good reliability (Nunnally and Bernstein 1994).

Table 3. Results for the Measurement Model

| Construct | Items | Standard loading | CR | AVE | Alpha |
|---|---|---|---|---|---|
| Performance Expectancy (PE) | PE1 | 0.75 | 0.74 | 0.69 | 0.72 |
| | PE2 | 0.73 | | | |
| | PE3 | 0.79 | | | |
| | PE4 | 0.69 | | | |
| Effort Expectancy (EE) | EE1 | 0.88 | 0.79 | 0.78 | 0.75 |
| | EE2 | 0.81 | | | |
| | EE3 | 0.74 | | | |
| | EE4 | 0.68 | | | |
| Social Influence (SI) | SI1 | 0.76 | 0.89 | 0.84 | 0.89 |
| | SI2 | 0.75 | | | |
| | SI3 | 0.68 | | | |
| | SI4 | 0.71 | | | |
| Facilitating Conditions (FC) | FC1 | 0.83 | 0.95 | 0.91 | 0.91 |
| | FC2 | 0.77 | | | |
| | FC3 | 0.78 | | | |

| Construct | Items | Standard loading | CR | AVE | Alpha |
|---|---|---|---|---|---|
| Website quality (WQ) | WQ1 | 0.87 | 0.91 | 0.66 | 0.90 |
|  | WQ2 | 0.77 |  |  |  |
|  | WQ3 | 0.89 |  |  |  |
|  | WQ4 | 0.86 |  |  |  |
|  | WQ5 | 0.78 |  |  |  |
| Usage behaviour (USE) | USE1 | 0.76 | 0.78 | 0.76 | 0.73 |
|  | USE2 | 0.83 |  |  |  |
|  | USE3 | 0.71 |  |  |  |
|  | USE4 | 0.82 |  |  |  |

To assess for discriminant validity, the square root of the average variance extracted (AVE) for each construct was compared with the inter-factor correlations between that construct and all other constructs. If the AVE is higher than the squared inter-scale correlations of the construct, then it shows good discriminant validity (Gefen et al. 2000; Hair et al. 2006). As shown in Table 4, for each factor, the square root of AVE is larger than the correlation coefficients with other factors and that confirms sufficient discriminant validity.

Table 4. Discriminant Validity Results for the Measurement Model

| Construct | PE | EE | SI | FC | WQ | USE |
|---|---|---|---|---|---|---|
| PE | 0.84 |  |  |  |  |  |
| EE | 0.22 | 0.89 |  |  |  |  |
| SI | 031 | 0.30 | 0.92 |  |  |  |
| FC | 0.27 | 0.50 | 0.25 | 0.95 |  |  |
| WQ | 0.33 | 0.05 | 0.06 | 0.44 | 0.95 |  |
| USE | 0.32 | 0.01 | 0.27 | 0.36 | 0.43 | 0.88 |

**Evaluation of the Structural Model**

As previously mentioned, the second step is to assess the structural model which includes the testing of the theoretical hypothesis and the relationships between latent constructs provided through the employed SEM technique and the use of AMOS software. Table 5 lists the path coefficients and their significance. Table 5 presents the path coefficients and their significance for each hypothesis, while the proposed structural model is depicted in Figure 3. Overall, the results of the proposed research model show a good fit: ($\chi^2$ = 605.35, $df$ = 341, $\chi^2/df$ = 1.775, GFI =0.910, TLI = 0.920, CFI = 0.930, IFI = 0.920, RMSEA = 0.071, SRMR= 0.071). Overall, four out of five hypotheses were supported by the data. All hypotheses (H1, H2, H4 and H5) representing the relationship among the main constructs (PE, EE, FC and WQ) to USE were supported in this study. The hypothesis that was not supported was H3: SI to USE. Social influence (SI) did not significantly predict usage behaviour of e-government services (-0.03, *n.s.*); therefore, H3 was not supported. As shown in Table 5, performance expectancy (PE) positively predicted usage behaviour (0.34, $p$ < 0.001); therefore, H1 was supported. Second, effort expectancy (EE) significantly predicted behavioural intent (0.39, $p$ < 0.001); therefore, H2 was supported. Third, social influence (SI) did not significantly predict behavioural intent (-0.03, *n.s.*); therefore, H3 was not supported. Fourth, facilitating conditions (FC) positively predicted usage behaviour (0.48, $p$ < 0.001), providing support for H4. Lastly, the website quality (WQ) construct in the e-government website positively predicted usage behaviour to use e-government services (0.74, $p$ < 0.001); thus H5 was supported.

Table 5. Structural Model Results

| Path (Hypothesis) | Standardised path coefficient (Beta) | t-value | Hypothesis testing result |
|---|---|---|---|
| PE →USE (H1) | 0.34 | 4.22*** | Supported |
| EE→ USE (H2) | 0.39 | 4.57 *** | Supported |
| SI→USE (H3) | -0.03 | 0.81 n.s. | Not supported |
| FC→USE(H4) | 0.48 | 3.20*** | Supported |
| WQ→USE(H5) | 0.74 | 4.92*** | Supported |

Model fit indices: $\chi^2$ = 605.35, $df$ = 341, $\chi^2/df$ = 1.775, GFI =0.910, TLI = 0.920, CFI = 0.930, IFI = 0.920, RMSEA = 0.071, SRMR= 0.071 *** $p < 0.001$; n.s. Not significant.

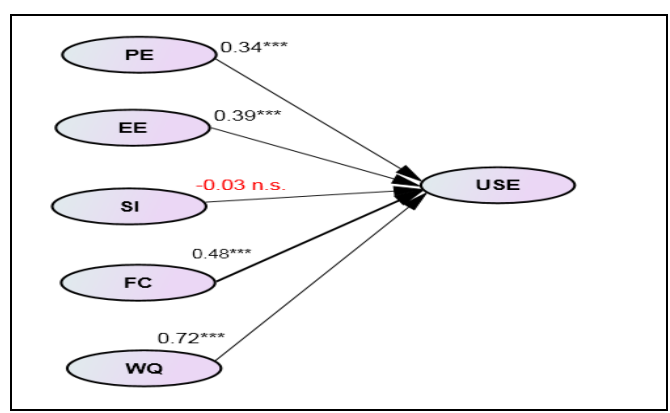

Figure 3: Structural model with standardized path coefficients

## DISCUSSION OF THE FINDINGS

The proposed research model UTAUT was empirically tested through a series of processes and steps to effectively carry out the research result and finding for quantitative data. This section will discuss the results and findings with respect to the variables in the proposed UTAUT research model: effort expectancy (EE), performance expectancy (PE), social influences (SI), website quality (WQ) and their relationship with the dependent variable usage behaviour (USE). The results of this study provide support for a majority of the study hypotheses proposed at the beginning of this study and elaborated as follows:

- *Performance Expectancy*
  In this study, performance expectancy (PE) is understood as the degree to which the user believes that using e-government services will facilitate communication with government in terms of benefits, saving time and money, improving the quality of government services and increasing equity between all citizens. The research result supports the hypothesis H1, which assumes that performance expectancy (PE) positively predicts behavioural intention (BI) to use e-government services. The effect of performance expectancy (PE) on usage behaviour (USE) was significant and strong and that definitely reflects the benefits obtained from using e-government services. This result was consistent with previous researches findings (Al-Qeisi 2009; Garfield 2005; Louho et al. 2006; Rosen 2005; and Venkatesh et al. 2003).

- *Effort Expectancy*
  The effort expectancy (EE) variable in this study is defined as the degree of ease associated with the use of e-government services system in KSA. It was measured by the perception of ease by which one could learn, use, and become skilful at using these systems. The link between effort expectancy (EE) and usage behaviour (USE) was significant and supported by the research finding (H2). Therefore, this finding was consistent with other study results which confirmed that effort expectancy has a strong effect on use intention (Birth and Irvine 2009; Helaiel 2009; Louho et al. 2006; Rosen 2005; Venkatesh et al. 2003).

- *Social Influence*
  The social influence (SI) construct was defined as the extent to which an individual perceives that it is important in the opinions of others that he or she should use e-government services. This was measured by the perception of how social communications will affect user's intention to use e-government services. The study result revealed the insignificant impact of social influence on usage behaviour (USE) of e-government services. As a result, the relationship and hypothesis (H3) between SI and USE was unsupported. This result confirms previous findings reported in several studies (Davis et al. 1989; Karahanna and Straub 1999; Rosen 2005; Taylor and Todd 1995a; Venkatesh and Davis 2000; Venkatesh et al. 2003). This result and previous findings indicate that the use of e-government systems was a personal and individual issue and not affected by social influence. Venkatesh et al. (2003) reported that the usage of a system depends on the user's beliefs, rather than on others' opinions or advice.

- *Facilitating Conditions*
  Facilitating conditions (FC) refers to the availability of technological and organizational resources used to support the use of the e-government system (Venkatesh et al. 2003). FC was measured by the perception of accessing required resources, and the necessary knowledge and technical support needed to use e-government services systems. The study result confirmed that the facilitating conditions (FC) have a direct and significant effect on usage behaviour (USE) of e-government services. That result supports the established direct link (H4) between facilitating conditions (FC) and usage behaviour (USE). This result was comparable with other empirical studies, including those by Hung et al. (2006), Taylor and Todd (1995b), and Venkatesh et al. (2003).

- *Website Quality*
  In this study, website quality (WQ) was integrated into the UTAUT model as an independent variable. It was measured based on several principles, including: technical quality, content quality, appearance quality, accessibility and availability. The study result confirmed that website quality (WQ) had a positive and significant effect on usage behaviour (USE) of e-government services. The relationship between website quality (WQ) and usage behaviour (USE) was the greatest and most powerful link with the standardized regression weights estimates. This result affirmed that the website quality factor directly impacts usage behaviour (USE) of e-government services in KSA; moreover, this impact is greater than any other construct in the proposed UTAUT model. This finding is consistent with the results of several studies, all of which reported that website quality affects behavioural intention, usage behaviour, and user satisfaction in their decision to adopt e-government systems (Ahn et al. 2007; DeLone and Mclean 2003; Nelson et al. 2005; Wixom and Todd, 2005). Therefore, this significant result demonstrates the success of adding website quality as an independent variable to the UTAUT model.

## CONCLUSION

The purpose of this study was to investigate the factors affecting the acceptance and use of e-government services in Saudi Arabia and, more specifically, to explore the role of website quality as an important factor on the adoption of e-government services. The finding of this study has explored a number of interesting findings. First, it confirms and affirms the need for dimensions of website quality (WQ) such as content quality, appearance quality, accessibility, ease of use, and good website design. Second, with respect to the main constructs of the UTAUT model, the finding showed that effort expectancy (EE), performance expectancy (PE), facilitating conditions (FC), and website quality (WQ) contribute significantly to citizen adoption of e-government services and directly affect the usage behaviour (USE) of e-government services. Third, the influence of the social influence (SI) variable on the usage behaviour (USE) of e-government services was insignificant for Saudi citizens. However, this study was conducted in the Kingdom of Saudi Arabia, so the analysis is based on the perception of the Saudi citizens. So, this finding confirms that online based transactions such e-government systems are essentially private and done alone with the computer, lowering the social influence factor on that process. The contribution of the paper is that it sucussed to integrate Website Quality construct with the UTAUT model to study the adoption of e-government services on KSA context. Further, this study is a significant step to identify and understand the obstacles of e-government services in KSA and then to provide and improve practical solutions to increase the adoption level. Moreover, the findings of this study provide an empirical result for other developing counties that have the same context with KSA and face the same difficulties for the adoption of e-government services (G2C) in their own country. However, this study has limitations too. First, it only focused on the main constructs of the UTAUT model and does not include the moderator's effect on the main relationships. If the moderator's relationships had been included, the analysis would have become complex and hard to maintain in a paper like this, given the space restriction. Second, we have studied only one type of e-government services (G2C), which aims to provide citizens with comprehensive electronic services and information. Thus, future research can build on our study model of the UTAUT in different countries as well as different types of e-government applications (especially B2C). Third, this study is based on quantitative data collected from 400 respondents, so this work could be extended using qualitative data to investigate more in-depth perceptions about other factors that affect the e-government services adoption. Furthermore, other variables such as trust and culture could be integrated in the UTAUT model to obtain a better and deeper understanding of e-government services adoption.

## APPENDIX

**MEASUREMENT SCALE AND ITEMS**

| **Performance expectancy** *(adapted from Venkatesh et al. 2003)* | |
|---|---|
| PE1 | Using e-government services enables me to accomplish my needs from public sectors more quickly and more efficiently. |
| PE2 | Using e-government services increases equity between all citizens. |
| PE3 | Using e-government services would save citizens' time. |
| PE4 | Using e-government services increases the quality of services. |
| **Effort Expectancy** *(adapted from Venkatesh et al. 2003)* | |
| EE1 | Learning to use the e-government services system is easy. |
| EE2 | Using e-government services system is easy. |
| EE3 | It is easy for me to become skilful at using the e-government services system. |
| EE4 | By using the e-government system I am able to obtain governmental services easily. |
| **Social Influence** *(adapted from Venkatesh et al. 2003)* | |
| SI1 | People who are important to me think that I should use e-government services. |
| SI2 | People who influence my behaviour think I should use e-government services. |
| SI3 | I would use e-government services if my friends and colleagues used them. |
| SI4 | Government sectors encourage citizens to use the e-government services system. |
| **Facilitating Conditions** *(adapted from Venkatesh et al. 2003)* | |
| FC1 | I have the resources necessary to use e-government services. |
| FC2 | I have the knowledge necessary to use e-government services. |
| FC3 | There is a specific person or group available for assistance with any technical problem I may encounter. |
| **Website Quality** *(adapted from Venkatesh et al. 2003)* | |
| WQ1 | Government websites appear safe and secure for carrying out transactions. |

| | | |
|---|---|---|
| WQ2 | Government websites look attractive and use fonts and colour properly. |
| WQ3 | Government websites look organized. |
| WQ4 | Government websites are always up and available 24/7. |
| WQ5 | Content of government websites is useful and updated regularly. |
| **Usage Behaviour of e-government service** (adapted from Aladwani 2006) | |
| USE1 | I really want to use e-government services to perform my governmental requests. |
| USE2 | I frequently use e-government services. |
| USE3 | I use e-government services on a regular basis. |
| USE4 | Most of my governmental requests are done through e-government services. |

## COPYRIGHT